# Sensitivity of the Power Spectra of Magnetization Fluctuations in Low Barrier Nanomagnets to Barrier Height Modulation and Defects

Md Ahsanul Abeed and Supriyo Bandyopadhyay[1]

*Department of Electrical and Computer Engineering, Virginia Commonwealth University*
*Richmond, VA 23284, USA*
*abeedma@vcu.edu, sbandy@vcu.edu*

Nanomagnets with small shape anisotropy energy barriers on the order of the thermal energy have unstable magnetization that fluctuates randomly in time. They have recently emerged as promising hardware platforms for stochastic computing and machine learning because the random magnetization states can be harnessed for probabilistic bits. Here, we have studied how the statistics of the magnetization fluctuations (e.g. the power spectral density) is affected by (i) moderate variations in the barrier height of the nanomagnet and (ii) the presence of structural defects, in order to assess how robust the computing platform is. We found that the power spectral density is relatively insensitive to moderate barrier height change and also relatively insensitive to the presence of small localized defects. However, extended (delocalized) defects, such as thickness variations over a significant fraction of the nanomagnet, affect the power spectral density very noticeably. As a result, small variations in the shape (causing small variations in the barrier height), or small localized defects, are relatively innocuous and tolerable, but significant variation of the nanomagnet thickness is not. Consequently, tight control over the nanomagnet thickness must be maintained for stochastic computing applications.

---

[1] Corresponding author

# 1. Introduction

There is a recent surge of interest in employing low energy barrier nanomagnets (LBM), with barrier heights on the order of the thermal energy, as hardware accelerators in stochastic computing [1-4]. In LBMs that have in-plane magnetic anisotropy, the magnetization points in random directions on the surface at different instants of time due to thermal noise, leading to random magnetization fluctuation. These random magnetization states can be utilized for probabilistic computing and machine learning [4]. For this, and all other kinds of probabilistic computing, it will be necessary for the *statistics* of the magnetization fluctuation to be relatively unaffected by unavoidable small variations in nanomagnet parameters such as shape (that would alter the energy barrier) or the presence of structural defects incurred during fabrication.

In this work, we show that the fluctuation statistics is indeed insensitive to moderate variations in the energy barrier height of the nanomagnet (due to small variations in shape) and also has weak sensitivity to small localized imperfections. However, extended defects, such as thickness variations over a significant fraction of the nanomagnet, have a very noticeable effect on the power spectral density, and hence the fluctuation statistics. Thus, it would be imperative to exercise tight control over the nanomagnet *thickness* for stochastic computing applications, more so than over the shape or small localized defects. That might be challenging if the nanomagnets are fabricated by metal evaporation on to a patterned resist or sputtering.

# 2. Simulations

In this work, we have studied thermally induced magnetization fluctuations in an LBM made of cobalt using the micromagnetic simulator MuMax3 [5]. The LBM that we considered is a thin elliptical disk of small eccentricity (nearly circular) with major axis dimension 100 nm, minor axis dimension 99.7 nm and thickness 6 nm, as shown in Fig. 1. The calculated shape anisotropy energy barrier in this nanomagnet is 1.3 kT at room temperature (300 K).

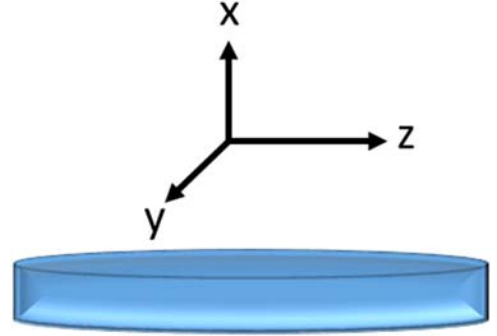

Fig. 1: A low barrier nanomagnet

We pick the *z*-axis along the major axis of the ellipse, the *y*-axis along the minor axis and the *x*-axis as perpendicular to the LBM's plane. We then start the MuMax3 simulation with the components

$$[m_z(t=0)=0.99;\ m_x(t=0)=0.141;\ m_y(t=0)=0]$$

and monitor the time variation in the presence of a random thermal field given by

$$\vec{H}_{th}(t) = H_x(t)\hat{x} + H_y(t)\hat{y} + H_z(t)\hat{z}$$

$$H_i(t) = \sqrt{\frac{2\alpha kT}{\gamma(1+\alpha^2)\mu_0 M_s \Omega \Delta t}} G_{0,1}(t) \qquad (i=x,y,z)$$

(1)

where $\alpha$ is the Gilbert damping factor of cobalt ($\alpha = 0.01$), $\gamma = 2\mu_B\mu_0/\hbar$, $\mu_B$ is the Bohr magneton, $\mu_0$ is the permeability of free space, $M_s$ is the saturation magnetization of cobalt ($1.1 \times 10^6$ A/m), $\Omega$ is the nanomagnet volume and $\Delta t$ is the time step used in the simulation (0.1 ps). We assumed that the exchange constant of cobalt is $3 \times 10^{-11}$ J/m. We ignore magneto-crystalline anisotropy, assuming that the nanomagnets are either amorphous or polycrystalline. All calculations are carried out for room temperature (300 K). The simulation is carried out for 1 ns, to calculate the auto-correlation function of the magnetization component fluctuation along the major axis, defined as

$$C(\tau) = \int_{-\infty}^{\infty} dt \left[ m_z(t) m_z(t+\tau) \right] \quad (2)$$

The simulations allow us to determine the magnetization component $m_z(t)$ at any instant of time $t$, and that allows us to calculate the auto-correlation function in Equation (2) within the window of 1 ns. The auto-correlation function would obviously have been different if we had chosen a different time window.

## 3. Results and Discussion

In Fig. 2, we show the calculated auto-correlation functions of two LBMs, having energy barriers of 1.3 kT and 2.2 kT, within the 1 ns window. The two nanomagnets are identical except that the latter's minor axis is 0.2 nm smaller. Note that even though the barrier heights differ by ~70%, the auto-correlation functions differ very slightly, showing that it is relatively insensitive to barrier height. The auto-correlation function is therefore robust against moderate variations in the barrier height and hence it is not imperative to control the shape of an LBM (which determines the barrier height) too tightly for stochastic computing.

According to the Wiener-Khinchin Theorem, the power spectral density of the magnetization fluctuation is the Fourier transform of the auto-correlation function:

$$S(f) = \int_{-\infty}^{\infty} d\tau C(t) e^{-i2\pi f \tau} \quad (3)$$

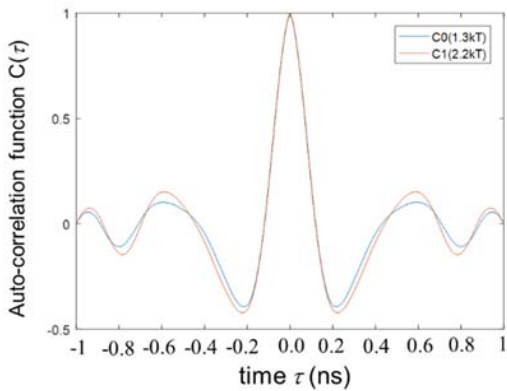

Fig. 2: Normalized auto-correlation functions of two LBMs with energy barriers of 1.3 kT and 2.2 kT computed within a time window of 1 ns.

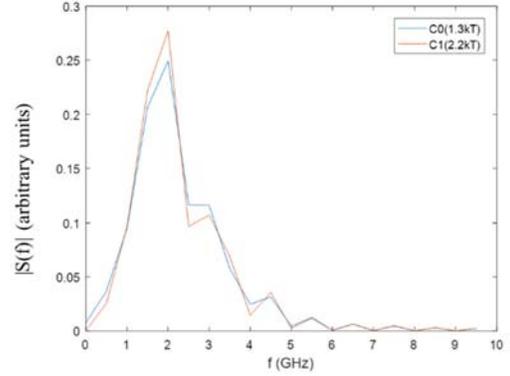

Fig. 3: Power spectral densities of the two LBMs with energy barriers of 1.3 kT and 2.2 kT computed from the data in Fig. 2.

In Fig. 3, we plot the power spectral densities of the magnetization fluctuations of the two LBMs with barrier heights of 1.3 kT and 2.2 kT, computed from the auto-correlation functions in Fig. 2. They are practically indistinguishable, showing that the power spectra of the magnetization fluctuations are not particularly sensitive to moderate variations in the energy barrier height. Note that this moderate variation accrues from a mere 0.2 nm variation in one dimension; hence, this kind of variation may be unavoidable.

Finally, we investigated if the power spectrum can be affected by the presence of structural defects in the nanomagnets. To this end, we considered four different types of structural defects depicted in the insets of Fig 4. They are commonplace in nanomagnets fabricated with e-beam lithography followed by metal evaporation and lift-off [6, 7]. These defects can be classified into "localized" (small hole or hillock in the center) or "extended" (two different thicknesses in two different halves of the nanomagnets, or a rim around the periphery). The power spectra do not change much in the presence of localized defects, but change very significantly when extended defects are present. This is shown in Fig. 4. Therefore, it would be more imperative to maintain tight control over LBM thickness rather than over the shape, or the presence of small localized defects.

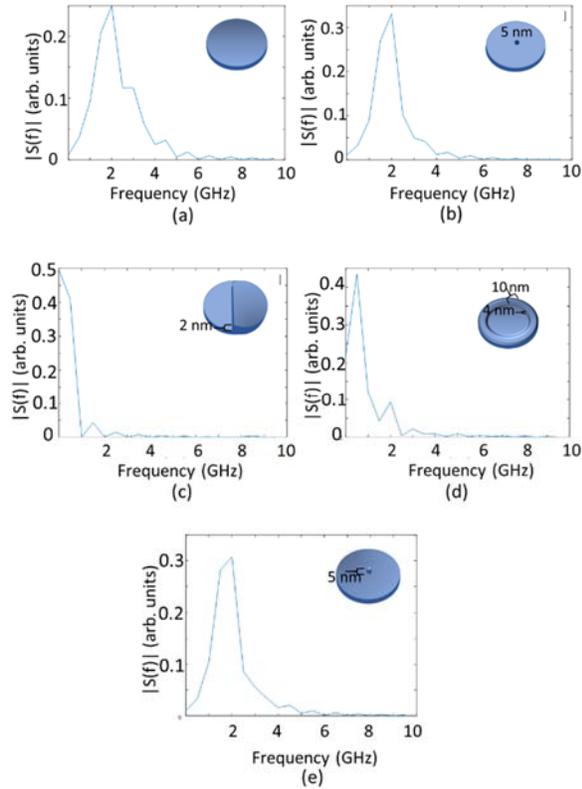

Fig. 4: Power spectral densities for: (a) defect-free nanomagnet, (b) a defective nanomagnet with a 5-nm diameter through hole in the center, (c) a defective nanomagnet where the thickness in one half is 6 nm and the other 8 nm, (d) a defective nanomagnet with an annulus around the periphery whose width is 10 nm and height 4 nm, and (e) a defective nanomagnet with a 5 nm diameter and 5 nm tall cylindrical hillock at the center.

## 4. Conclusions

In conclusion, we have shown that the statistical parameters of magnetization fluctuation in low barrier nanomagnets are relatively insensitive to moderate variations in the barrier height, or to the presence of small localized defects. However, the statistical parameters are quite sensitive to significant thickness variations across the surface of the nanomagnets. In many applications, such as binary stochastic neurons used as hardware accelerators for machine learning [4], parameters such as the correlation time (defined as the fill-width-at-half-maximum of the auto-correlation function) are important and they could become uncontrollable if the nanomagnet thickness variation could not be constrained to a small value.